\documentclass[english]{article}
\usepackage[T1]{fontenc}
\usepackage[utf8]{inputenc}
\usepackage{geometry}
\usepackage{array}
\usepackage{multirow}
\usepackage{amsmath}
\usepackage{amssymb}
\usepackage{graphicx}
\usepackage{setspace}

\makeatletter

\providecommand{\tabularnewline}{\\}

\usepackage{jheppub}

\author[a,b]{Tamas Gombor}
\affiliation[a]{MTA-ELTE “Momentum” Integrable Quantum Dynamics Research Group, Department of Theoretical Physics, Eötvös
  Loránd University}
\affiliation[b]{Holographic QFT Group, Wigner Research Centre for Physics, Budapest, Hungary}

\emailAdd{gombort@caesar.elte.hu}
\abstract{We study the integrable boundaries and crosscaps of classical sigma models.
We show that there exists a classical analog of the integrability condition and KT-relation of
 the boundary and crosscap states of quantum spin chains. 
We also classify the integrable crosscaps for various sigma models including examples which are relevant in
the AdS/CFT correspondence at strong coupling.}

\makeatother

\usepackage{babel}
\begin{document}
\title{Integrable crosscaps in classical sigma models}
\maketitle

\section{Introduction}

In recent years, intensive research has been done in the defect versions
of AdS/CFT duality. The probe brane defects in the string theory side
were introduced in \cite{Karch:2001cw} and in the gauge theory side
they correspond to defect conformal field theories (dCFTs) \cite{DeWolfe:2001pq}.
For theories with domain wall defects, the one-point functions at
weak coupling can be obtained from an overlap between a \emph{boundary
state} (corresponding to the defect) and the Bethe states (corresponding
to the single trace operators) of a spin chain (describing the scalar
sector of the CFT) \cite{deLeeuw:2015hxa,Buhl-Mortensen:2015gfd,Kristjansen:2021abc}.
Similar overlaps appear also for the three-point functions which contain
one single trace and two determinant type operators \cite{Jiang:2019xdz,Jiang:2019zig,Yang:2021hrl}.

It was possible to write these overlaps in closed forms \cite{deLeeuw:2016umh,DeLeeuw:2018cal,DeLeeuw:2019ohp,Gombor:2022aqj}.
These closed forms were validated only by numerical checks. The reason
of the existence such exact overlaps is the underlying integrability
of the boundary states. The definition of the integrability for boundary
states of spin chains was developed in \cite{Piroli:2017sei} (based
on \cite{Ghoshal:1993tm}). We call a boundary state integrable if
it preserves the half of the conserved charges of the spin chain.
This condition can be written in a compact form 
\begin{equation}
\langle B|\mathcal{T}(u)=\langle B|\mathcal{T}^{\beta}(-u),\label{eq:Qint}
\end{equation}
where $\langle B|$ is the boundary state and $\mathcal{T}(u),\mathcal{T}^{\beta}(u)$
are transfer matrices of the spin chain which are connected by a $\mathbb{Z}_{2}$
automorphism $\beta$ of the underlying symmetry algebra\footnote{Typically, $\beta$ is the identity or the charge conjugation.}.
In \cite{Gombor:2021hmj} a systematic algebraic method was introduced
which proves the previously proposed form of the exact overlaps for
a wide class of boundary states. The heart of this derivation is the
so-called $KT$-relation
\begin{equation}
K_{0}(u)\langle B|T_{0}(u)=\langle B|T_{0}^{\beta}(-u)K_{0}(u),\label{eq:QKT}
\end{equation}
where $T(u),T^{\beta}(u)$ are the monodromy matrices and $K(u)$
is the $K$-matrix which acts only on the auxiliary space.

Recently, an other type of states called \emph{crosscap states} \cite{Caetano:2021dbh}
was introduced for spin chains which are the analogous versions of
the crosscap states of 2d CFT \cite{Ishibashi:1988kg}. The difference
between the boundary and the crosscap states are the following. While
the boundary states identify the neighboring sites of the spin chain,
the crosscap identifies the antipodal sites. In \cite{Gombor:2022deb}
the $KT$-relation was generalized for crosscap states and it was
used to classify the crosscap states for all $\mathfrak{gl}(N)$ symmetric
spin chains. The previously proposed formula of \cite{Caetano:2021dbh}
for overlaps was also proved for a wide class of crosscap states based
on the $KT$-relation. In \cite{Caetano:2022mus} it was argued that
integrable crosscap states are appeared in the one-point functions
of the $\mathcal{N}=4$ SYM on the $\mathbb{R}\mathrm{P}^{4}$ spacetime.

As we already mentioned, at strong coupling the defect corresponds
to a probe D-brane which can be describe as boundary conditions of
the string sigma models. For certain boundary conditions the integrability
has been already shown \cite{Dekel:2011ja,Linardopoulos:2021rfq,Linardopoulos:2022wol}.
During these derivations the boundaries were putted on space therefore
it was showed that infinity many conserved charges exist on the worldsheet
of the \emph{open strings} which are attached on the D-brane. However,
in the holographic description of the one-point function we have a
closed string (corresponding to the operator) which is annihilated
on the D-brane. In this setup the boundary is in time, and intuitively,
the integrability means that the boundary condition preserves the
half of the worldsheet conserved charges of the \emph{closed strings}.
It is clear that it would be a classical analog of the quantum integrability
condition (\ref{eq:Qint}).

The goal of this paper is twofold. Firstly, we want to show that classical
analogs of the integrability conditions (\ref{eq:Qint}) and $KT$-relations
(\ref{eq:QKT}) exist for classical sigma models with boundaries in
time. We also show that these relations are automatically satisfied
for the classical reflection matrices of \cite{Dekel:2011ja,Linardopoulos:2021rfq,Linardopoulos:2022wol}.
We also generalize the integrability condition and $KT$-relations
for the crosscaps of classical sigma models. The second goal is to
classify the integrable crosscaps of the sigma models which appear
in the AdS/CFT at strong coupling. This paper is not intended to provide
a complete holographic description, we will not go beyond the classic
sigma model. However, this classification for the sigma models can
be a good starting point for more comprehensive future investigations.

The organization of the paper is as follows. In section \ref{sec:Crosscaps-of-sigma}
we give the definition of the crosscaps of sigma models. In section
\ref{sec:Lax-description} we review the Lax description (Lax-connection,
transfer matrix etc.) of the sigma models. In section \ref{sec:Boundaries-in-time}
we derive the integrability condition and $KT$-relation when the
boundaries are in time. In section \ref{sec:Lax-descriptions-of}
we generalize the integrability condition for crosscaps and classify
them for the sigma models with target spaces $SU(N)$, $S^{N-1}$,
$AdS_{N-1}$ and $\mathbb{C}\mathrm{P}^{N-1}$. In section \ref{sec:Crosscaps-in-AdS/CFT}
we classify the integrable crosscaps of the sigma models which appear
in the $AdS_{5}/CFT_{4}$ and $AdS_{4}/CFT_{3}$ dualities at strong
coupling. 

\section{Crosscaps of sigma models\label{sec:Crosscaps-of-sigma}}

In this section we define the crosscaps for 2d sigma models. Let $\mathcal{M}$
be a semi-Riemann manifold (target space) with local coordinates $X^{I}$
where $I=1,\dots,\dim\mathcal{M}$ and metric $G_{IJ}(X)$. We also
define a 2d manifold (worldsheet) with local coordinates $\sigma^{\mu}=\left(\sigma^{0},\sigma^{1}\right)=\left(\tau,\sigma\right)$
and the worldsheet metric $\eta$. The fields in the 2d sigma model
are maps between the worldsheet and the target space: $X:\Sigma\to\mathcal{M}$.
The dynamics are given by the action
\begin{equation}
S[X]=\int_{\Sigma}d\tau d\sigma G_{IJ}(X)\partial_{\mu}X^{I}\partial^{\mu}X^{J}.
\end{equation}
Let $\alpha$ be a $\mathbb{Z}_{2}$ isometry of $\mathcal{M}$ which
acts on the fields as $\alpha:X\to X^{\alpha}$. It is clear that
if we have a solution $X^{I}(\tau,\sigma)$ of the equations of motion
then the transformed fields $X^{\alpha,I}(\tau,\sigma)$ also satisfy
them. 

At first let us consider an infinite cylinder as worksheet: $\Sigma=\mathbb{R}\times S^{1}$.
We choose the local coordinates as $\tau\in\mathbb{R}$ and $\sigma\in\left[0,L\right)$
with the identification $\sigma+L\equiv\sigma$. We define the crosscap
by a restriction of the allowed configurations of the fields. We allow
only the configurations which are invariant under the following $\mathbb{Z}_{2}$
transformation
\begin{equation}
X(\tau,\sigma)=X^{\alpha}(-\tau,\sigma+\frac{L}{2}).\label{eq:id}
\end{equation}
We call (\ref{eq:id}) as \emph{crosscap identification} and the field
configurations which satisfy (\ref{eq:id}) are the \emph{crosscap
configurations}. Let us divide the worldsheet into two regions $\Sigma=\Sigma_{+}\cup\Sigma_{-}$
where $\Sigma_{+}=\left[0,\infty\right)\times S^{1}$ and $\Sigma_{-}=\left(-\infty,0\right]\times S^{1}$.
For the crosscap configurations it is enough to give the fields on
$\Sigma_{-}$ since the crosscap identification (\ref{eq:id}) gives
also the fields on $\Sigma_{+}$. It is clear that we can choose any
configuration on almost the full $\Sigma_{-}$. We get non-trivial
conditions only on the intersection $\Sigma_{+}\cap\Sigma_{-}$ which
is the $\tau=0$ circle. Substituting to (\ref{eq:id}) we obtain
that
\begin{equation}
X^{I}\Biggr|_{\tau=0,\sigma=\sigma_{0}}=X^{\alpha,I}\Biggr|_{\tau=0,\sigma=\sigma_{0}+\frac{L}{2}}.\label{eq:Crosscap1}
\end{equation}
We have an other non--trivial smoothness condition for the time derivatives
\begin{equation}
\left(\partial_{\tau}X^{I}\right)\Biggr|_{\tau=0,\sigma=\sigma_{0}}=-\left(\partial_{\tau}X^{\alpha,I}\right)\Biggr|_{\tau=0,\sigma=\sigma_{0}+\frac{L}{2}}.\label{eq:Crosscap2}
\end{equation}
Choosing any field configuration on $\Sigma_{-}$ with conditions
(\ref{eq:Crosscap1}) and (\ref{eq:Crosscap2}) we can uniquely extend
it to a crosscap configuration on the full $\Sigma$. In the sections
\ref{sec:Boundaries-in-time}, \ref{sec:Lax-descriptions-of} and
\ref{sec:Crosscaps-in-AdS/CFT} we concentrate on sigma models on
the worldsheet $\Sigma_{-}$ and we call the conditions (\ref{eq:Crosscap1})
and (\ref{eq:Crosscap2}) as \emph{crosscap conditions}.

\section{Lax description\label{sec:Lax-description}}

Let us consider an integrable 2 dimensional field theory on the worldsheet
$\Sigma$ and a Lax connection $A(\lambda)=A_{\tau}(\lambda)d\tau+A_{\sigma}(\lambda)d\sigma\in\Omega_{1}(\Sigma)\otimes\mathfrak{gl}(N)$
where $\Omega_{1}(\Sigma)$ are the one-forms on $\Sigma$ and $\lambda\in\mathbb{C}$
is the spectral parameter. The Lax connection satisfies the zero curvature
equation
\begin{equation}
\mathrm{d}A(\lambda)+A(\lambda)\wedge A(\lambda)=0.\label{eq:zeroc}
\end{equation}
This equation is the local manifestation of the path independence
of the holonomies of the Lax connection
\begin{equation}
\left[\overleftarrow{\mathcal{P}\exp}\int_{\gamma_{1}}-A(\lambda)\right]=\left[\overleftarrow{\mathcal{P}\exp}\int_{\gamma_{2}}-A(\lambda)\right],\label{eq:bulkInt}
\end{equation}
where $\gamma_{1}$ and $\gamma_{2}$ are homotopic curves with same
end points. The equation (\ref{eq:bulkInt}) is the real heart of
the integrability since it guaranties the existence of infinite many
integrals of motion. We can define monodromy matrices
\begin{equation}
T(\lambda|\tau)=\overleftarrow{\mathcal{P}\exp}\int_{\sigma=0}^{L}-A_{\sigma}(\lambda)d\sigma.\label{eq:mon}
\end{equation}
Using the integrability condition (\ref{eq:bulkInt}) we obtain that
the time evolution of the monodromy matrix can be written as
\begin{equation}
T(\lambda|\tau_{2})=U_{L}(\tau_{2},\tau_{1})T(\lambda|\tau_{1})U_{0}(\tau_{2},\tau_{1})^{-1},
\end{equation}
where 
\begin{equation}
U_{\sigma}(\tau_{2},\tau_{1})=\overleftarrow{\mathcal{P}\exp}\int_{\tau=\tau_{1}}^{\tau_{2}}-A_{\tau}(\lambda)d\tau.
\end{equation}
For the periodic boundary condition $\sigma\equiv\sigma+L$ we obtain
that
\begin{equation}
U_{L}(\tau_{2},\tau_{1})=U_{0}(\tau_{2},\tau_{1}),
\end{equation}
therefore the time evolution of the monodromy matrix is a similarity
transformation
\begin{equation}
T(\lambda|\tau_{2})=U_{0}(\tau_{2},\tau_{1})T(\lambda|\tau_{1})U_{0}(\tau_{2},\tau_{1})^{-1}.
\end{equation}
The trace of the monodromy matrix (transfer matrix)
\begin{equation}
\mathcal{T}(\lambda|\tau)=\mathrm{Tr}T(\lambda|\tau)\label{eq:transfer}
\end{equation}
generates the conserved quantities
\begin{equation}
\mathcal{T}(\lambda):=\mathcal{T}(\lambda|\tau_{1})=\mathcal{T}(\lambda|\tau_{2}).
\end{equation}

\subsection*{Examples}

In the following we define four well known example for the integrable
sigma model. For more details see the review \cite{Zarembo:2017muf}.

\subsubsection*{Principal chiral model}

Let $G$ and $\mathfrak{g}$ be a Lie group and the corresponding
Lie algebra. Defining the target space as $g(\tau,\sigma)\in G$,
the current $J=g^{-1}dg\in\Omega_{1}(\Sigma)\otimes\mathfrak{g}$
is a Lie-algebra valued one-form. The equation of motions
\begin{equation}
d*J=0,\quad dJ+J\wedge J=0\label{eq:eom}
\end{equation}
 are equivalent to the zero curvature equation (\ref{eq:zeroc}) of
the Lax connection
\begin{equation}
A(\lambda)=\frac{1}{1+\lambda^{2}}J+\frac{\lambda}{1+\lambda^{2}}*J,\label{eq:Lax}
\end{equation}
where $*$ is the Hodge duality.

\subsubsection*{$O(N)$ sigma model}

For the $O(N)$ sigma model the target space is the $N-1$ dimensional
sphere $S^{N-1}$ with radius $1$. Let us parameterize this sphere
with the coordinates $\phi_{i}(\tau,\sigma)\in\mathbb{R}$ for which
$\sum_{i}\phi_{i}\phi_{i}=\mathbf{\boldsymbol{\phi}}^{t}\boldsymbol{\phi}=1$
($\boldsymbol{\phi}^{t}=\left(\phi_{1},\dots,\phi_{N}\right)$ is
a row vector and $\boldsymbol{\phi}$ is column vector). The equations
of motion are
\begin{equation}
\partial^{2}\boldsymbol{\phi}+\boldsymbol{\phi}\left(\partial_{\mu}\boldsymbol{\phi}^{t}\partial^{\mu}\boldsymbol{\phi}\right)=0.
\end{equation}
We can introduce a group element $h(\tau,\sigma)\in O(N)$ as
\begin{equation}
h=\mathbf{1}-2\boldsymbol{\phi}\boldsymbol{\phi}^{t},
\end{equation}
for which
\begin{equation}
h^{2}=(\mathbf{1}-2\boldsymbol{\phi}\boldsymbol{\phi}^{t})(\mathbf{1}-2\boldsymbol{\phi}\boldsymbol{\phi}^{t})=\mathbf{1},
\end{equation}
therefore $h^{-1}=h=h^{t}$. Introducing the current 
\begin{equation}
J=hdh=2\boldsymbol{\phi}d\boldsymbol{\phi}^{t}-2d\boldsymbol{\phi}\boldsymbol{\phi}^{t},
\end{equation}
the equations of motion have the form (\ref{eq:eom}) therefore the
Lax-connection can be written as (\ref{eq:Lax}).

\subsubsection*{Sigma model on the $AdS_{N}$}

The target space is the $N$ dimensional anti de-Sitter $AdS_{N}$
with radius $1$. Let us parameterize this space with the coordinates
$X_{i}(\tau,\sigma)\in\mathbb{R}$ where $i=-1,0,1,\dots,N-1$ for
which 
\begin{equation}
-\left(X_{-1}\right)^{2}-\left(X_{0}\right)^{2}+\sum_{i=1}^{N-1}\left(X_{i}\right)^{2}=\sum_{i,j=-1}^{N-1}\eta_{i,j}X_{i}X_{j}=\mathbf{X}^{t}\eta\mathbf{X}=-1,
\end{equation}
where $\mathbf{X}^{t}=(X_{-1},X_{0},\dots,X_{N-1})$ is a row vector
and
\begin{equation}
\eta=\mathrm{diag}(-1,-1,1,1,\dots,1).
\end{equation}
 The equation of motion is
\begin{equation}
\partial^{2}\mathbf{X}-\mathbf{X}\left(\partial_{\mu}\mathbf{X}^{t}\eta\partial^{\mu}\mathbf{X}\right)=0.
\end{equation}
 We can introduce a group element $h(\tau,\sigma)\in O(2,N-1)$ as
\begin{equation}
h=\eta-2\mathbf{X}\mathbf{X}^{t},
\end{equation}
for which $h=h^{t}$ and
\begin{equation}
h\eta h=(\eta+2\mathbf{X}\mathbf{X}^{t})\eta(\eta+2\mathbf{X}\mathbf{X}^{t})=\eta,
\end{equation}
therefore $h^{-1}=\eta h\eta$. Introducing the current 
\begin{equation}
J=h^{-1}dh=2\eta(d\mathbf{X}\mathbf{X}^{t}-\mathbf{X}d\mathbf{X}^{t}),
\end{equation}
the equations of motion are (\ref{eq:eom}) therefore the Lax-connection
has the form (\ref{eq:Lax}). We will also use the Poincaré coordinates
\begin{align}
X_{-1} & =\frac{z}{2}\left(1+\frac{1+x^{2}}{z^{2}}\right),\\
X_{N-1} & =\frac{z}{2}\left(1-\frac{1-x^{2}}{z^{2}}\right),\\
X_{i} & =\frac{x_{i}}{z},\qquad i=0,1,\dots,N-2,
\end{align}
where $x^{2}=-\left(x_{0}\right)^{2}+\sum_{i=1}^{N-2}\left(x_{i}\right)^{2}$.

\subsubsection*{Sigma model on the $\mathbb{C}\mathrm{P}^{N-1}$}

For the $\mathbb{C}\mathrm{P}^{N-1}$ sigma model we use the action
\begin{equation}
S=\int d\tau d\sigma\left(\mathcal{D}_{\mu}\mathbf{Y}\right)^{\dagger}\left(\mathcal{D}^{\mu}\mathbf{Y}\right),
\end{equation}
where $\mathbf{Y}\in\mathbb{C}^{N}$ with the constraint $\mathbf{Y}^{\dagger}\mathbf{Y}=1$
and we introduced the covariant derivative as
\begin{equation}
\mathcal{D}_{\mu}=\partial_{\mu}-i\mathcal{A}_{\mu}.
\end{equation}
where $\mathcal{A}_{\mu}$ is a $U(1)$ gauge field. The equations
of motion are
\begin{align}
0 & =\mathcal{D}_{\mu}\mathcal{D}^{\mu}\mathbf{Y}+\left(\mathcal{D}_{\mu}\mathbf{Y}\right)^{\dagger}\left(\mathcal{D}^{\mu}\mathbf{Y}\right)\mathbf{Y},\\
\mathcal{A}_{\mu} & =i\left(\partial_{\mu}\mathbf{Y}^{\dagger}\right)\mathbf{Y}=-i\mathbf{Y}^{\dagger}\left(\partial_{\mu}\mathbf{Y}\right).
\end{align}
 We can introduce a group element $h(\tau,\sigma)\in U(N)$ as
\begin{equation}
h=\mathbf{1}-2\mathbf{Y}\mathbf{Y}^{\dagger},
\end{equation}
for which
\begin{equation}
h^{2}=(\mathbf{1}-2\mathbf{Y}\mathbf{Y}^{\dagger})(\mathbf{1}-2\mathbf{Y}\mathbf{Y}^{\dagger})=\mathbf{1},
\end{equation}
therefore $h^{-1}=h=h^{\dagger}$. Introducing the current 
\begin{equation}
J=hdh=2\mathbf{Y}d\mathbf{Y}^{\dagger}-2d\mathbf{Y}\mathbf{Y}^{\dagger}+4i\mathcal{A}\mathbf{Y}\mathbf{Y}^{\dagger},
\end{equation}
the equations of motion are (\ref{eq:eom}) therefore the Lax-connection
has the form (\ref{eq:Lax}).

\section{Boundaries in time\label{sec:Boundaries-in-time}}

We make a detour in this section. Before we move on to the Lax description
of the crosscaps, we first consider the case where the usual boundary
condition is placed not in space but in time. The integrable boundary
conditions in space (the boundary is the $\sigma=0$ line) have been
already analyzed in several papers, e.g. \cite{MacKay:2001bh,MacKay:2004rz,MacKay:2011zs,Aniceto:2017jor,Gombor:2018ppd}.
In these situations the so-called double row transfer matrices generate
the conserved charges on the half plane or the strip. In the following
we analyze what happens when we put the same integrable boundaries
to the $\tau=0$ circle of $\Sigma_{-}$.

\begin{figure}
\begin{centering}
\includegraphics[width=0.6\textwidth]{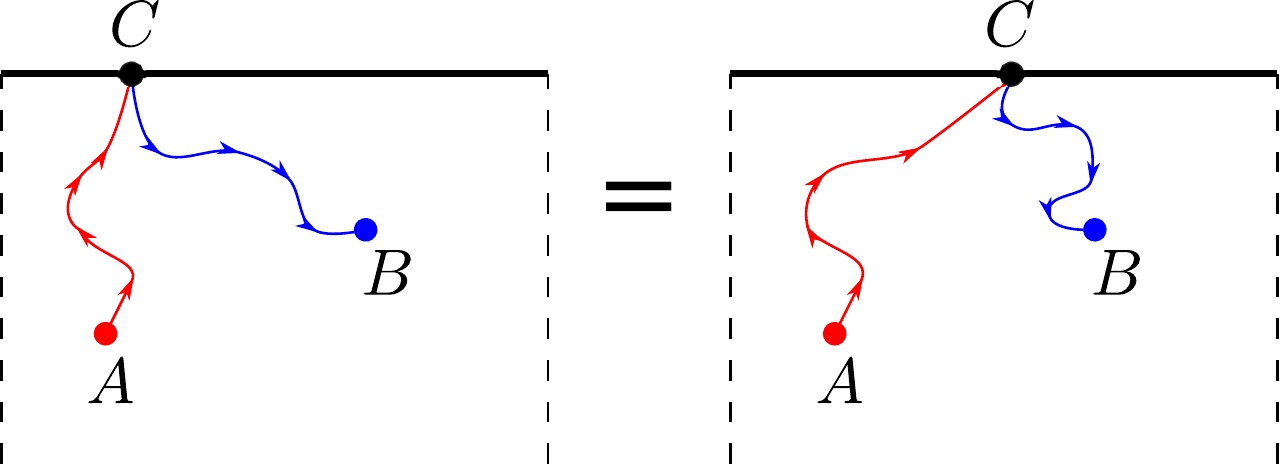}
\par\end{centering}
\caption{Path independent holonomy. The red curve corresponds to the holonomy
of $A(\lambda)$ between $A$ and $C$. The blue curve corresponds
to the holonomy of $A^{\beta}(-\lambda)$ between $C$ and $B$.}

\label{fig:boundaryBC}
\end{figure}

We saw that the phenomena of integrability comes from path independent
holonomies. In the boundary case we can define holonomies which attach
to the boundary 
\begin{equation}
\left[\overleftarrow{\mathcal{P}\exp}\int_{\gamma_{2}}-A^{\beta}(-\lambda)\right]\kappa(\lambda)\left[\overleftarrow{\mathcal{P}\exp}\int_{\gamma_{1}}-A(\lambda)\right],\label{eq:holB}
\end{equation}
where $\kappa$ is a reflection matrix and $\alpha$ is an automorphism
which leaves the flatness condition invariant i.e.
\begin{equation}
\mathrm{d}A^{\beta}(\lambda)+A^{\beta}(\lambda)\wedge A^{\beta}(\lambda)=0.
\end{equation}
The endpoints of the curves $\gamma_{1},\gamma_{2}$ are $[C,A],[B,C]$
where $A$ and $B$ are fixed and $C$ is a common point which is
on the boundary. According to the flatness condition of the Lax-connection
we can freely deform the curves $\gamma_{1},\gamma_{2},$ and it is
natural to say, the boundary condition is integrable if the holonomy
(\ref{eq:holB}) is independent from the common point $C$, see figure
\ref{fig:boundaryBC}. 

We can differentiate this condition and we obtain the following constraint
for the Lax connection
\begin{equation}
\kappa(\lambda)A_{\sigma}(\lambda)-A_{\sigma}^{\beta}(-\lambda)\kappa(\lambda)\Biggr|_{\tau=0}=\partial_{\sigma}\kappa(\lambda).\label{eq:kL}
\end{equation}
This is the usual equation of the integrable boundaries, only the
space and time coordinates are replaced. Some solutions which are
relevant in the AdS/CFT correspondence can be found in \cite{Dekel:2011ja,Linardopoulos:2021rfq,Linardopoulos:2022wol}.
\begin{figure}
\begin{centering}
\includegraphics[width=0.6\textwidth]{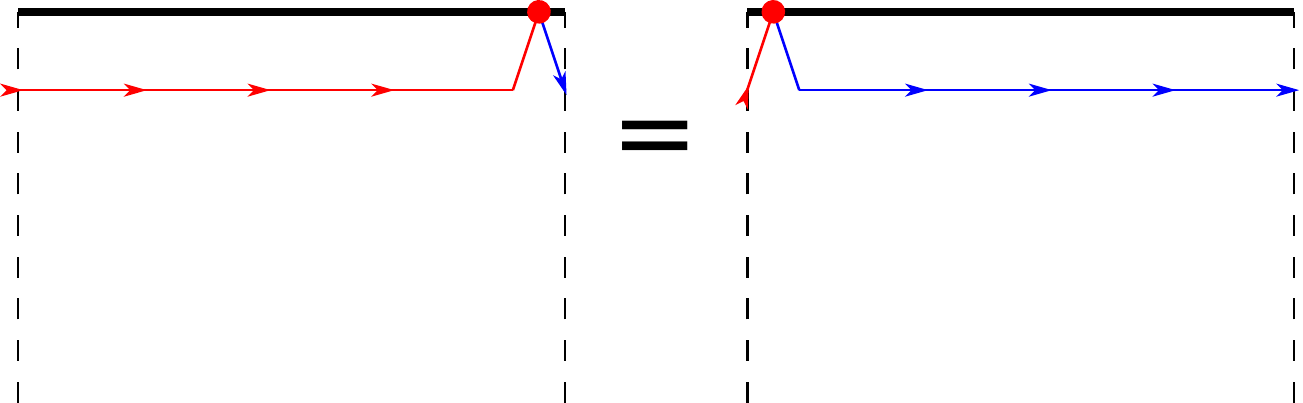}
\par\end{centering}
\caption{The $KT$-equation. The red curve corresponds to the holonomy of $A(\lambda)$
and the blue curve corresponds to the holonomy of $A^{\beta}(-\lambda)$.}

\label{fig:boundary}
\end{figure}

Using the above defined boundary flatness condition, we can obtain
the following equation (see figure \ref{fig:boundary})
\begin{equation}
\kappa(\lambda)T(\lambda)=T^{\beta}(-\lambda)\kappa(\lambda),\label{eq:KT}
\end{equation}
where $T(\lambda)$ is the monodromy matrix (\ref{eq:mon}) at $\tau=0$.
The equation (\ref{eq:KT}) is the classical analog of the quantum
$KT$-relation (\ref{eq:QKT}) which was introduced in \cite{Gombor:2021uxz}.
At this point we define two $\beta$-transformations.
\begin{enumerate}
\item $A^{\beta}=A$ (identity).
\item $A^{\beta}=-A^{t}$ (charge conjugation). 
\end{enumerate}
For the first case the $KT$-relation is simplified as
\begin{equation}
\kappa(\lambda)T(\lambda)=T(-\lambda)\kappa(\lambda),
\end{equation}
which is the classical analog of the untwisted $KT$-relation of \cite{Gombor:2021hmj}.
For the second case let us introduce a new notation for the $\beta$-transformed
monodromy matrix
\begin{equation}
\widehat{T}(\lambda):=T^{\beta}(\lambda)=\overleftarrow{\mathcal{P}\exp}\int_{\sigma=0}^{L}A_{\sigma}^{t}(\lambda)d\sigma.
\end{equation}
We can see that
\begin{equation}
T(\lambda)\widehat{T}^{t}(\lambda)=1.\label{eq:inv}
\end{equation}
In the second case (where $A^{\beta}=-A^{t}$), the $KT$-relation
(\ref{eq:KT}) is equivalent to
\begin{equation}
\kappa(\lambda)T(\lambda)=\widehat{T}(-\lambda)\kappa(\lambda),
\end{equation}
which is the classical analog of the twisted $KT$-relation of \cite{Gombor:2021hmj}. 

From the $KT$-relation and the definition of the transfer matrix
(\ref{eq:transfer}) we can easily derive the condition
\begin{equation}
\mathcal{T}(\lambda)=\mathcal{T}^{\beta}(-\lambda),
\end{equation}
therefore the half of the charges are vanishing which is the classical
analog of integrability condition of boundary states \cite{Piroli:2017sei}.
Specifying the $\beta$-transformation we obtain the following conditions
\begin{align}
\mathcal{T}(\lambda) & =\mathcal{T}(-\lambda),\\
\mathcal{T}(\lambda) & =\widehat{\mathcal{T}}(-\lambda),
\end{align}
which are the classical analog of the untwisted and twisted quantum
integrability conditions \cite{Gombor:2020kgu}. 

It is worth to mention some related works. The classsical $KT$-relation
(\ref{eq:KT}) with some minor modifications were already appeared
in \cite{Caudrelier:2014oia}. This paper investigated the non-linear
Schrödinger equation and introduced a dual, equal-space, Poisson bracket
which describes a Hamiltonian flow in the space direction. In this
dual Hamiltonian flow a usual defect in space can be considered as
a defect in time. The Liouville integrability was also proved in this
dual picture. These ideas have been developed along various directions
and for various key models \cite{Caudrelier:2014gsa,Avan:2015gja,Doikou:2016oej,Doikou:2019njk}.

\section{Lax description of crosscaps\label{sec:Lax-descriptions-of}}

In this section we can generalize the argument of the previous section
for crosscap conditions (\ref{eq:Crosscap1}),(\ref{eq:Crosscap2}). 

\begin{figure}
\begin{centering}
\includegraphics[width=0.6\textwidth]{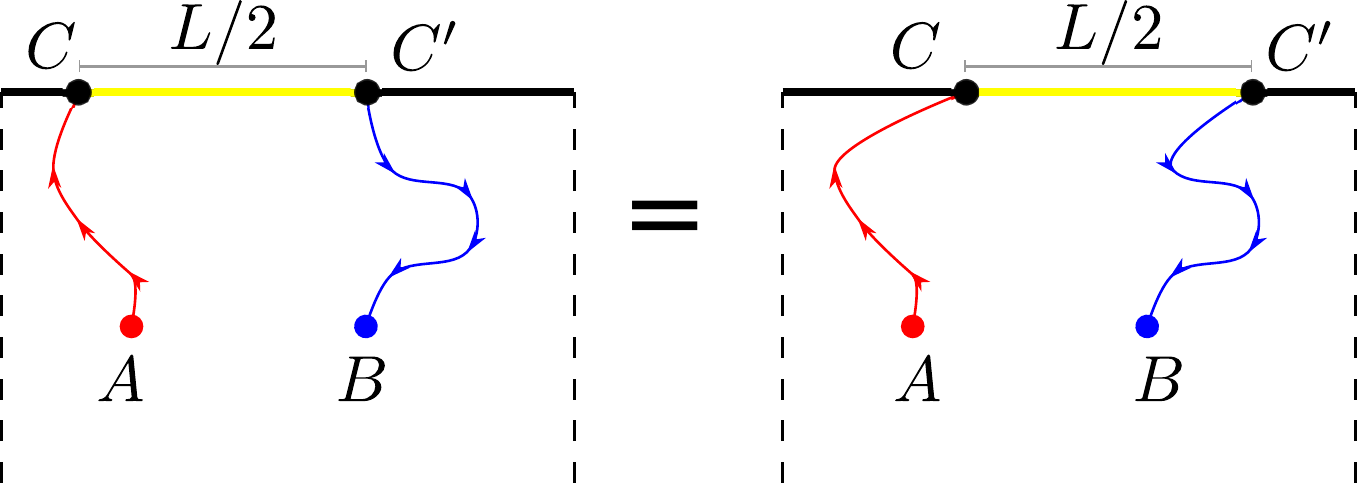}
\par\end{centering}
\caption{Path independent holonomy.}

\label{fig:crosscapBC}
\end{figure}

Now we identify the antipodal points of the time boundary $\tau=0$
therefore the flatness condition has to be modified in the following
way: the holonomy
\begin{equation}
\left[\overleftarrow{\mathcal{P}\exp}\int_{\gamma_{2}}-A^{\beta}(-\lambda)\right]\kappa\left[\overleftarrow{\mathcal{P}\exp}\int_{\gamma_{1}}-A(\lambda)\right]\label{eq:crossB}
\end{equation}
is independent of the path, see figure \ref{fig:crosscapBC}. The
endpoints of the curves $\gamma_{1},\gamma_{2}$ are $[C,A],[B,C']$
where $A$ and $B$ are fixed and $C=(0,\sigma_{0}),C'=(0,\sigma_{0}+\frac{L}{2})$.
The holonomy (\ref{eq:crossB}) is independent from the space coordinate
$\sigma_{0}$.

The local version of this flatness condition is \footnote{We concentrate on the constant $\kappa$-matrices}
\begin{equation}
\kappa A_{\sigma}(\lambda)\Biggr|_{\tau=0,\sigma=\sigma_{0}}=A_{\sigma}^{\beta}(-\lambda)\kappa\Biggr|_{\tau=0,\sigma=\sigma_{0}+\frac{L}{2}}.\label{eq:kL-1}
\end{equation}
As in the previous section, we distinguish two types of automorphism
$\beta$:
\begin{enumerate}
\item $A^{\beta}=A$ (identity).
\item $A^{\beta}=-A^{t}$ (charge conjugation). 
\end{enumerate}
Let us start with the first case when $\beta=\mathrm{id}$. Applying
the crosscap condition twice, we obtain that
\begin{equation}
A_{\sigma}(\lambda)\Biggr|_{\tau=0,\sigma=\sigma_{0}}=\kappa^{-1}A_{\sigma}(-\lambda)\kappa\Biggr|_{\tau=0,\sigma=\sigma_{0}+\frac{L}{2}}=\left(\kappa^{-1}\right)^{2}A_{\sigma}(\lambda)\kappa^{2}\Biggr|_{\tau=0,\sigma=\sigma_{0}},
\end{equation}
where we used the periodic boundary condition $\sigma\equiv\sigma+L$.
Since we do not want to introduce any local constraint for the fields,
the crosscap is consistent if
\begin{equation}
\kappa^{2}=c\mathbf{1},\label{eq:contUTW}
\end{equation}
where $c\in\mathbb{C}$. Since the flatness condition is linear in
$\kappa$ we can always choose a normalization where $\kappa^{2}=\mathbf{1}$.
For the second case ($\beta$ is the charge conjugation) the crosscap
condition reads as
\begin{equation}
\kappa A_{\sigma}(\lambda)\Biggr|_{\tau=0,\sigma=\sigma_{0}}=-A_{\sigma}^{t}(-\lambda)\kappa\Biggr|_{\tau=0,\sigma=\sigma_{0}+\frac{L}{2}}.
\end{equation}
Applying this crosscap condition twice, we obtain that
\begin{equation}
A_{\sigma}(\lambda)\Biggr|_{\tau=0,\sigma=\sigma_{0}}=-\kappa^{-1}A_{\sigma}^{t}(-\lambda)\kappa\Biggr|_{\tau=0,\sigma=\sigma_{0}+\frac{L}{2}}=\kappa^{-1}\kappa^{t}A_{\sigma}(\lambda)(\kappa^{-1})^{t}\kappa\Biggr|_{\tau=0,\sigma=\sigma_{0}}.
\end{equation}
Similarly as in the previous case, we do not want to introduce any
local constraint for the fields therefore the crosscap is consistent
if
\begin{equation}
\kappa^{t}=\pm\kappa.\label{eq:constTW}
\end{equation}
\begin{figure}
\begin{centering}
\includegraphics[width=0.6\textwidth]{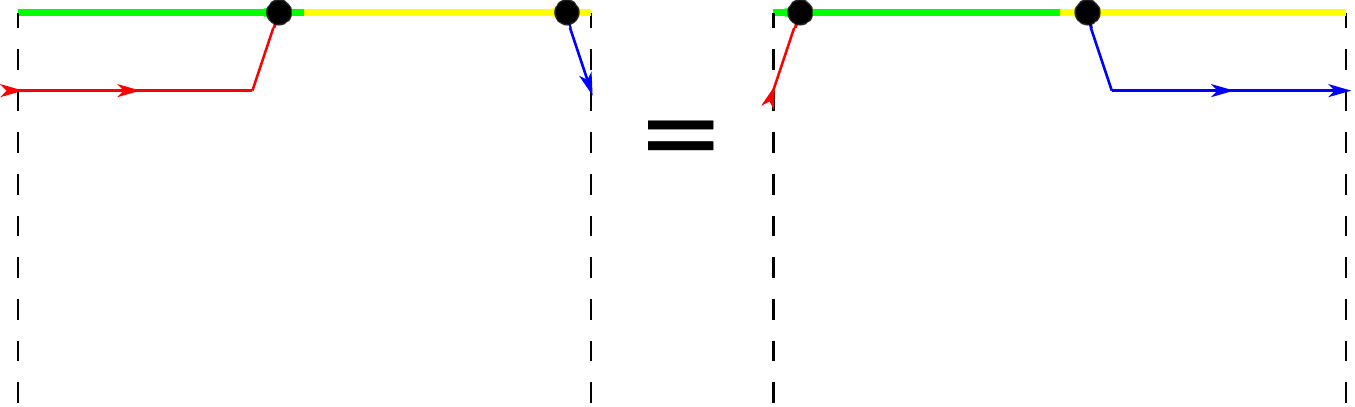}
\par\end{centering}
\caption{The $KT$-equation.}

\label{fig:crosscap}
\end{figure}

Using the global (\ref{eq:crossB}) or the local (\ref{eq:kL-1})
crosscap condition of the Lax connection we can obtain the following
equation (see figure \ref{fig:crosscap})
\begin{equation}
\kappa T^{(1)}(\lambda)=T^{(2),\alpha}(-\lambda)\kappa,\label{eq:KT-1}
\end{equation}
where $T^{(1)}(\lambda),T^{(2)}(\lambda)$ are two new monodromy matrices
at $\tau=0:$
\begin{align}
T^{(1)}(\lambda) & =\overleftarrow{\mathcal{P}\exp}\int_{0}^{L/2}-A_{\sigma}(\lambda|\tau=0,\sigma)d\sigma.\\
T^{(2)}(\lambda) & =\overleftarrow{\mathcal{P}\exp}\int_{L/2}^{L}-A_{\sigma}(\lambda|\tau=0,\sigma)d\sigma.
\end{align}
It is obvious that the full monodromy matrix is the product of them:
\begin{equation}
T(\lambda)=T^{(2)}(\lambda)T^{(1)}(\lambda).
\end{equation}
The equation (\ref{eq:KT-1}) is the classical analog of the quantum
$KT$-relation which was introduced in \cite{Gombor:2022deb}. For
the first case ($\beta$ is the identity) the $KT$-relation is simplified
as
\begin{equation}
\kappa T^{(1)}(\lambda)=T^{(2)}(-\lambda)\kappa,
\end{equation}
which is the classical analog of the untwisted $KT$-relation of \cite{Gombor:2022deb}.
For the second case (where $A^{\beta}=-A^{t}$) the equation (\ref{eq:KT-1})
is equivalent to
\begin{equation}
\kappa T^{(1)}(\lambda)=\widehat{T}^{(2)}(-\lambda)\kappa,
\end{equation}
which is the classical analog of the twisted $KT$-relation of \cite{Gombor:2022deb}.
Using the connection (\ref{eq:inv}) between the monodromy matrices,
we can obtain an equivalent form of the twisted $KT$-relation
\begin{equation}
\kappa^{t}T^{(2)}(\lambda)=\widehat{T}^{(1)}(-\lambda)\kappa^{t}.
\end{equation}

Applying the untwisted $KT$-relation to the transfer matrix we obtain
that
\begin{equation}
\mathcal{T}(\lambda)=\mathrm{Tr}T^{(2)}(\lambda)T^{(1)}(\lambda)=\mathrm{Tr}\kappa T^{(1)}(-\lambda)\kappa^{-1}\kappa^{-1}T^{(2)}(-\lambda)\kappa.
\end{equation}
Using the constraint (\ref{eq:contUTW}) for the untwisted $\kappa$-matrix,
we just obtained that
\begin{equation}
\mathcal{T}(\lambda)=\mathcal{T}(-\lambda),
\end{equation}
which is the classical analog of the untwisted quantum integrability
condition \cite{Gombor:2022deb}.

Applying the twisted $KT$-relation to the transfer matrix we obtain
that
\begin{equation}
\mathcal{T}(\lambda)=\mathrm{Tr}T^{(2)}(\lambda)T^{(1)}(\lambda)=\mathrm{Tr}\left((\kappa^{t})^{-1}\widehat{T}^{(1)}(-\lambda)\kappa^{t}\kappa^{-1}\widehat{T}^{(2)}(-\lambda)\kappa\right).
\end{equation}
Using the constraint (\ref{eq:constTW}) for the twisted $\kappa$-matrix,
we just obtained that
\begin{equation}
\mathcal{T}(\lambda)=\widehat{\mathcal{T}}(-\lambda),
\end{equation}
which is the classical analog of the twisted quantum integrability
condition \cite{Gombor:2022deb}.

Let us analyze the solutions of the crosscap conditions for the models
which are described by the type of Lax pairs (\ref{eq:Lax}). The
crosscap conditions for the currents are
\begin{align}
J_{\sigma}\Biggr|_{\tau=0,\sigma=\sigma_{0}} & =\kappa^{-1}J_{\sigma}^{\beta}\kappa\Biggr|_{\tau=0,\sigma=\sigma_{0}+\frac{L}{2}}=J_{\sigma}^{\alpha}\Biggr|_{\tau=0,\sigma=\sigma_{0}+\frac{L}{2}}\label{eq:eqs}\\
J_{\tau}\Biggr|_{\tau=0,\sigma=\sigma_{0}} & =-\kappa^{-1}J_{\tau}^{\beta}\kappa\Biggr|_{\tau=0,\sigma=\sigma_{0}+\frac{L}{2}}=-J_{\tau}^{\alpha}\Biggr|_{\tau=0,\sigma=\sigma_{0}+\frac{L}{2}}\label{eq:eqt}
\end{align}
where we introduced the transformation
\begin{equation}
X^{\alpha}=\Omega(X):=\kappa^{-1}X^{\beta}\kappa.
\end{equation}
We already saw that $\Omega^{2}=\mathrm{id}$ for the consistent crosscaps.
We can see that the conditions (\ref{eq:eqs}),(\ref{eq:eqt}) are
equivalent to the crosscap conditions (\ref{eq:Crosscap1}),(\ref{eq:Crosscap2})
therefore the construction of this section indeed describes crosscaps. 

Let us decompose the current as $J_{\mu}=J_{\mu}^{(0)}+J_{\mu}^{(1)}$
where $\Omega(J_{\mu}^{(0)})=J_{\mu}^{(0)}$ and $\Omega(J_{\mu}^{(1)})=-J_{\mu}^{(1)}$.
Using these notations the crosscap condition simplifies as
\begin{align}
J_{\sigma}^{(0)}\Biggr|_{\tau=0,\sigma=\sigma_{0}} & =J_{\sigma}^{(0)}\Biggr|_{\tau=0,\sigma=\sigma_{0}+\frac{L}{2}}, & J_{\sigma}^{(1)}\Biggr|_{\tau=0,\sigma=\sigma_{0}} & =-J_{\sigma}^{(1)}\Biggr|_{\tau=0,\sigma=\sigma_{0}+\frac{L}{2}},\\
J_{\tau}^{(1)}\Biggr|_{\tau=0,\sigma=\sigma_{0}} & =J_{\tau}^{(1)}\Biggr|_{\tau=0,\sigma=\sigma_{0}+\frac{L}{2}}, & J_{\tau}^{(0)}\Biggr|_{\tau=0,\sigma=\sigma_{0}} & =-J_{\tau}^{(0)}\Biggr|_{\tau=0,\sigma=\sigma_{0}+\frac{L}{2}}.
\end{align}

\subsection*{Examples}

In the following we analyze the crosscap equations (\ref{eq:eqs}),(\ref{eq:eqt})
for some concrete sigma models which were introduced in section \ref{sec:Lax-description}.
More concretely, we specify the manifestation of these crosscap equations
on the concrete parametrization of the target spaces. Since we already
show that the conditions (\ref{eq:eqs}),(\ref{eq:eqt}) are equivalent
to the crosscap conditions (\ref{eq:Crosscap1}),(\ref{eq:Crosscap2})
we concentrate on the explicit forms of the $\alpha$-isometries and
the corresponding residual symmetries.

\subsubsection*{Principal chiral fields}

Let us concentrate on the $SU(N)$ principal chiral field i.e. $\mathfrak{g}=\mathfrak{su}(N)$.
For the $\beta=\mathrm{id}$ case we have $\kappa^{2}=\mathbf{1}.$
Using the global $SU(N)$ symmetry we can diagonalize the $\kappa$-s
therefore we have
\begin{equation}
\kappa=\mathrm{diag}(\underbrace{1,\dots,1}_{k},\underbrace{-1,\dots,-1}_{N-k}).
\end{equation}
The $\alpha$-automorphism acts on the fields as $g^{\alpha}=\kappa g\kappa^{-1}$
and the residual symmetry is $SU(k)\times SU(N-k)\times U(1)$. In
the second case ($\beta$ is the charge conjugation) $g^{\alpha}=\kappa\left(g^{t}\right)^{-1}\kappa^{-1}$
and the residual symmetries are $SO(N)$ or $Sp(N)$ for $\kappa=\kappa^{t}$
or $\kappa=-\kappa^{t}$, respectively. 

This classification of classical crosscaps are completely analog with
the classification of the crosscap states of the quantum $\mathfrak{su}(N)$
symmetric spin chains \cite{Gombor:2022deb}.

\subsubsection*{$O(N)$ sigma model\label{subsec:-sigma-model}}

For the $O(N)$ sigma model the current is an element of $\mathfrak{so}(N)$
i.e. $-J^{t}=J$ therefore only the $\beta=\mathrm{id}$ case is relevant.
Since the current is anti-symmetric the consistent crosscaps require
both conditions $\kappa^{2}=c\mathbf{1}$ and $\kappa^{t}=\pm\kappa$
therefore we have two classes of $\kappa$-matrices (up to global
$O(N)$ rotations)
\begin{align}
\kappa & =\mathrm{diag}(\underbrace{1,\dots,1}_{k},\underbrace{-1,\dots,-1}_{N-k}),\\
\kappa & =\left(\begin{array}{cc}
0 & \mathbf{1}_{\frac{N}{2}\times\frac{N}{2}}\\
-\mathbf{1}_{\frac{N}{2}\times\frac{N}{2}} & 0
\end{array}\right).
\end{align}
The symmetric case has residual symmetry $SO(k)\times SO(N-k)$ and
the anti-symmetric one has $U(\frac{N}{2})$. In the symmetric case
the solution of the equation (\ref{eq:eqs}) is 
\begin{align}
\phi_{i}^{\alpha} & =\phi_{i},\qquad i=1,\dots,k,\label{eq:ONC1}\\
\phi_{i}^{\alpha} & =-\phi_{i},\qquad i=k+1,\dots,N.\label{eq:ONC2}
\end{align}
For the anti-symmetric $\kappa$, the equation (\ref{eq:eqs}) has
the solution
\begin{align}
\phi_{i}^{\alpha} & =-\phi_{\frac{N}{2}+i},\qquad i=1,\dots,\frac{N}{2},\\
\phi_{\frac{N}{2}+i}^{\alpha} & =\phi_{i},\qquad i=\frac{N}{2}+1,\dots,N.
\end{align}
We can see that this is not a consistent crosscap because the square
of this transformation is $-1$.

In summary, for the $O(N)$ sigma model the integrable crosscaps correspond
to the residual symmetries $SO(k)\times SO(N-k)$ and the concrete
identifications are given by the equations (\ref{eq:ONC1}-\ref{eq:ONC2}).
In the table \ref{tab:S5} we enumerate these possibilities explicitly
for the $O(6)$ model.
\begin{table}
\caption{Crosscaps of $O(6)$ sigma model.}

\begin{doublespace}
\begin{centering}
\begin{tabular}{|c|c|}
\hline 
Residual symmetry & $(\phi_{1},\phi_{2},\phi_{3},\phi_{4},\phi_{5},\phi_{6})^{\alpha}$\tabularnewline
\hline 
\hline 
\multirow{2}{*}{$SO(6)$} & $(\phi_{1},\phi_{2},\phi_{3},\phi_{4},\phi_{5},\phi_{6})$\tabularnewline
\cline{2-2} 
 & $(-\phi_{1},-\phi_{2},-\phi_{3},-\phi_{4},-\phi_{5},-\phi_{6})$\tabularnewline
\hline 
\multirow{2}{*}{$SO(5)$} & $(\phi_{1},\phi_{2},\phi_{3},\phi_{4},\phi_{5},-\phi_{6})$\tabularnewline
\cline{2-2} 
 & $(\phi_{1},-\phi_{2},-\phi_{3},-\phi_{4},-\phi_{5},-\phi_{6})$\tabularnewline
\hline 
\multirow{2}{*}{$SO(4)\times SO(2)$} & $(\phi_{1},\phi_{2},\phi_{3},\phi_{4},-\phi_{5},-\phi_{6})$\tabularnewline
\cline{2-2} 
 & $(\phi_{1},\phi_{2},-\phi_{3},-\phi_{4},-\phi_{5},-\phi_{6})$\tabularnewline
\hline 
$SO(3)\times SO(3)$ & $(\phi_{1},\phi_{2},\phi_{3},-\phi_{4},-\phi_{5},-\phi_{6})$\tabularnewline
\hline 
\end{tabular}
\par\end{centering}
\end{doublespace}
\label{tab:S5}
\end{table}

\subsubsection*{Sigma model on the $AdS_{N}$\label{subsec:Sigma-model-on}}

For the sigma model on the $AdS_{N}$ the current is an element of
$\mathfrak{so}(2,N-1)$ i.e. $-J^{t}=\eta J\eta$ therefore only the
$\beta=\mathrm{id}$ case is relevant for which we have the condition
$\kappa^{2}=c\mathbf{1}$. From the crosscap equation (\ref{eq:eqs})
we also obtain an other constraint since $\kappa J\kappa^{-1}\in\mathfrak{so}(2,N-1)$
i.e., $\kappa^{t}=\pm\eta\kappa\eta$. Let us start with the symmetric
$\kappa$-matrix
\begin{equation}
\kappa=\mathrm{diag}(1,1,\underbrace{1,\dots,1}_{N-2},-1).\label{eq:k1}
\end{equation}
This $\kappa$-matrix has $SO(2,N-2)$ symmetry. The equation (\ref{eq:eqs})
has the solution
\begin{equation}
\begin{split}X_{i}^{\alpha} & =X_{i},\qquad i=-1,\dots,N-2,\\
X_{N-1}^{\alpha} & =-X_{N-1}.
\end{split}
\end{equation}
Using the Poincaré coordinates we obtain that
\begin{equation}
\begin{split}x_{i}^{\alpha} & =\frac{x_{i}}{x^{2}+z^{2}},\qquad i=0,\dots,N-2,\\
z^{\alpha} & =\frac{z}{x^{2}+z^{2}}.
\end{split}
\label{eq:cross1}
\end{equation}
Using global $SO(2,N-1)$ isometries we can rotate the $\kappa$-matrix
to
\begin{equation}
\kappa=\mathrm{diag}(1,1,\underbrace{1,\dots,1}_{N-3},-1,1).\label{eq:k2}
\end{equation}
Which leads to the crosscap conditions
\begin{equation}
\begin{split}x_{i}^{\alpha} & =x_{i},\qquad i=0,\dots,N-3,\\
x_{N-2}^{\alpha} & =-x_{N-2},\\
z^{\alpha} & =z.
\end{split}
\label{eq:cross2}
\end{equation}
Since the $\kappa$-matrices $\eqref{eq:k1}$ and $\eqref{eq:k2}$
are connected by an isometry of $AdS_{N}$ the crosscaps (\ref{eq:cross1})
and (\ref{eq:cross2}) are also equivalent up to an isometry. 

The possible $\kappa^{2}=c\mathbf{1}$ matrices (up to global $SO(2,N-1)$
rotations) are
\begin{align}
\kappa & =\mathrm{diag}(1,1,\underbrace{1,\dots,1}_{k-1},\underbrace{-1,\dots,-1}_{N-k}),\\
\kappa & =\mathrm{diag}(1,-1,\underbrace{1,\dots,1}_{k-1},\underbrace{-1,\dots,-1}_{N-k}),\label{eq:secK}
\end{align}
where the first one has symmetry $SO(2,k-1)\times SO(N-k)$ and the
second one has $SO(1,k-1)\times SO(1,N-k)$. We can also get anti-symmetric
$\kappa$ but it leads to inconsistent crosscap just as for the $O(N)$
sigma model. In the first case the equation (\ref{eq:eqs}) has the
solution
\begin{align}
X_{i}^{\alpha} & =X_{i},\qquad i=-1,\dots,k-2,\label{eq:ONC1-1}\\
X_{i}^{\alpha} & =-X_{i},\qquad i=k-1,\dots,N-1,\label{eq:ONC2-1}
\end{align}
Using the Poincaré coordinates we obtain that
\begin{equation}
\begin{split}x_{i}^{\alpha} & =\frac{x_{i}}{x^{2}+z^{2}},\qquad i=0,\dots,k-2,\\
x_{i}^{\alpha} & =\frac{-x_{i}}{x^{2}+z^{2}},\qquad i=k-1,\dots,N-2,\\
z^{\alpha} & =\frac{z}{x^{2}+z^{2}}.
\end{split}
\label{eq:cross2-1}
\end{equation}
In the boundary of $AdS$ ($z=0$) these conditions are equivalent
to
\begin{equation}
\begin{split}x_{i}^{\alpha} & =\frac{x_{i}}{x^{2}},\qquad i=0,\dots,k-2,\\
x_{i}^{\alpha} & =\frac{-x_{i}}{x^{2}},\qquad i=k-1,\dots,N-2.
\end{split}
\end{equation}
For the second $\kappa$-matrix (\ref{eq:secK}), we can obtain similar
conditions and we show the result only in the Poincaré coordinates
\begin{equation}
\begin{split}x_{0}^{\alpha} & =\frac{-x_{0}}{x^{2}+z^{2}},\\
x_{i}^{\alpha} & =\frac{x_{i}}{x^{2}+z^{2}},\qquad i=1,\dots,k-2,\\
x_{i}^{\alpha} & =\frac{-x_{i}}{x^{2}+z^{2}},\qquad i=k-1,\dots,N-2,\\
z^{\alpha} & =\frac{z}{x^{2}+z^{2}}.
\end{split}
\label{eq:cross2-1-1}
\end{equation}
In the boundary of $AdS$ ($z=0$) these conditions are equivalent
to
\begin{equation}
\begin{split}x_{0}^{\alpha} & =\frac{-x_{0}}{x^{2}},\\
x_{i}^{\alpha} & =\frac{x_{i}}{x^{2}},\qquad i=1,\dots,k-2,\\
x_{i}^{\alpha} & =\frac{-x_{i}}{x^{2}},\qquad i=k-1,\dots,N-2.
\end{split}
\end{equation}

In summary, for the sigma model on the $AdS_{N}$ the integrable crosscaps
correspond to the residual symmetries $SO(2,k-1)\times SO(N-k)$ or
$SO(1,k-1)\times SO(1,N-k)$. In the tables \ref{tab:AdS5} and \ref{tab:AdS4}
we enumerate these possibilities explicitly for the $AdS_{5}$ and
$AdS_{4}$.

\begin{table}
\caption{Crosscaps of the sigma model on $AdS_{5}$.}

\begin{doublespace}
\begin{centering}
\begin{tabular}{|c|c|}
\hline 
Residual symmetry & $(x_{0},x_{1},x_{2},x_{3},z)^{\alpha}$\tabularnewline
\hline 
\hline 
\multirow{1}{*}{$SO(2,4)$} & $(x_{0},x_{1},x_{2},x_{3},z)$\tabularnewline
\hline 
\hline 
\multirow{1}{*}{$SO(2,3)$} & $\left(\frac{x_{0}}{x^{2}+z^{2}},\frac{x_{1}}{x^{2}+z^{2}},\frac{x_{2}}{x^{2}+z^{2}},\frac{x_{3}}{x^{2}+z^{2}},\frac{z}{x^{2}+z^{2}}\right)$\tabularnewline
\hline 
\multirow{1}{*}{$SO(2,2)\times SO(2)$} & $\left(\frac{x_{0}}{x^{2}+z^{2}},\frac{x_{1}}{x^{2}+z^{2}},\frac{x_{2}}{x^{2}+z^{2}},\frac{-x_{3}}{x^{2}+z^{2}},\frac{z}{x^{2}+z^{2}}\right)$\tabularnewline
\hline 
$SO(2,1)\times SO(3)$ & $\left(\frac{x_{0}}{x^{2}+z^{2}},\frac{x_{1}}{x^{2}+z^{2}},\frac{-x_{2}}{x^{2}+z^{2}},\frac{-x_{3}}{x^{2}+z^{2}},\frac{z}{x^{2}+z^{2}}\right)$\tabularnewline
\hline 
$SO(2)\times SO(4)$ & $\left(\frac{x_{0}}{x^{2}+z^{2}},\frac{-x_{1}}{x^{2}+z^{2}},\frac{-x_{2}}{x^{2}+z^{2}},\frac{-x_{3}}{x^{2}+z^{2}},\frac{z}{x^{2}+z^{2}}\right)$\tabularnewline
\hline 
\hline 
$SO(1,4)$ & $\left(\frac{-x_{0}}{x^{2}+z^{2}},\frac{-x_{1}}{x^{2}+z^{2}},\frac{-x_{2}}{x^{2}+z^{2}},\frac{-x_{3}}{x^{2}+z^{2}},\frac{z}{x^{2}+z^{2}}\right)$\tabularnewline
\hline 
\multirow{2}{*}{$SO(1,1)\times SO(1,3)$} & $\left(\frac{-x_{0}}{x^{2}+z^{2}},\frac{x_{1}}{x^{2}+z^{2}},\frac{-x_{2}}{x^{2}+z^{2}},\frac{-x_{3}}{x^{2}+z^{2}},\frac{z}{x^{2}+z^{2}}\right)$\tabularnewline
\cline{2-2} 
 & $\left(\frac{-x_{0}}{x^{2}+z^{2}},\frac{x_{1}}{x^{2}+z^{2}},\frac{x_{2}}{x^{2}+z^{2}},\frac{x_{3}}{x^{2}+z^{2}},\frac{z}{x^{2}+z^{2}}\right)$\tabularnewline
\hline 
$SO(1,2)\times SO(1,2)$ & $\left(\frac{-x_{0}}{x^{2}+z^{2}},\frac{x_{1}}{x^{2}+z^{2}},\frac{x_{2}}{x^{2}+z^{2}},\frac{-x_{3}}{x^{2}+z^{2}},\frac{z}{x^{2}+z^{2}}\right)$\tabularnewline
\hline 
\end{tabular}
\par\end{centering}
\end{doublespace}
\label{tab:AdS5}
\end{table}
\begin{table}
\caption{Crosscaps of the sigma model on $AdS_{4}$.}

\begin{doublespace}
\begin{centering}
\begin{tabular}{|c|c|}
\hline 
Residual symmetry & $(x_{0},x_{1},x_{2},z)^{\alpha}$\tabularnewline
\hline 
\hline 
\multirow{1}{*}{$SO(2,3)$} & $(x_{0},x_{1},x_{2},z)$\tabularnewline
\hline 
\hline 
\multirow{1}{*}{$SO(2,2)$} & $\left(\frac{x_{0}}{x^{2}+z^{2}},\frac{x_{1}}{x^{2}+z^{2}},\frac{x_{2}}{x^{2}+z^{2}},\frac{z}{x^{2}+z^{2}}\right)$\tabularnewline
\hline 
\multirow{1}{*}{$SO(2,1)\times SO(2)$} & $\left(\frac{x_{0}}{x^{2}+z^{2}},\frac{x_{1}}{x^{2}+z^{2}},\frac{-x_{2}}{x^{2}+z^{2}},\frac{z}{x^{2}+z^{2}}\right)$\tabularnewline
\hline 
$SO(2)\times SO(3)$ & $\left(\frac{x_{0}}{x^{2}+z^{2}},\frac{-x_{1}}{x^{2}+z^{2}},\frac{-x_{2}}{x^{2}+z^{2}},\frac{z}{x^{2}+z^{2}}\right)$\tabularnewline
\hline 
\hline 
$SO(1,3)$ & $\left(\frac{-x_{0}}{x^{2}+z^{2}},\frac{-x_{1}}{x^{2}+z^{2}},\frac{-x_{2}}{x^{2}+z^{2}},\frac{z}{x^{2}+z^{2}}\right)$\tabularnewline
\hline 
\multirow{2}{*}{$SO(1,1)\times SO(1,2)$} & $\left(\frac{-x_{0}}{x^{2}+z^{2}},\frac{x_{1}}{x^{2}+z^{2}},\frac{-x_{2}}{x^{2}+z^{2}},\frac{z}{x^{2}+z^{2}}\right)$\tabularnewline
\cline{2-2} 
 & $\left(\frac{-x_{0}}{x^{2}+z^{2}},\frac{x_{1}}{x^{2}+z^{2}},\frac{x_{2}}{x^{2}+z^{2}},\frac{z}{x^{2}+z^{2}}\right)$\tabularnewline
\hline 
\end{tabular}
\par\end{centering}
\end{doublespace}
\label{tab:AdS4}
\end{table}

\subsubsection*{Sigma model on the $\mathbb{C}\mathrm{P}^{N-1}$}

For the $\mathbb{C}\mathrm{P}^{N-1}$ sigma model the current is an
element of $\mathfrak{su}(N)$. For the $\beta=\mathrm{id}$ case
we have $\kappa^{2}=\mathbf{1}.$ Using the global $SU(N)$ symmetry
we can diagonalize the $\kappa$-s therefore we have
\begin{equation}
\kappa=\mathrm{diag}(\underbrace{1,\dots,1}_{k},\underbrace{-1,\dots,-1}_{N-k}).
\end{equation}
This $\kappa$-matrix has residual symmetry $SU(k)\times SU(N-k)\times U(1)$.
The solution of the equation (\ref{eq:eqs}) is 
\begin{align}
Y_{i}^{\alpha} & =Y_{i},\qquad i=1,\dots,k,\label{eq:ONC1-2}\\
Y_{i}^{\alpha} & =-Y_{i},\qquad i=k+1,\dots,N.\label{eq:ONC2-2}
\end{align}
In the second case (the charge conjugation case when $\kappa^{t}=\pm\kappa$)
we have two types of $\kappa$-matrices (up to global $SU(N)$ rotations)
\begin{align}
\kappa & =\mathbf{1},\\
\kappa & =\left(\begin{array}{cc}
0 & \mathbf{1}_{\frac{N}{2}\times\frac{N}{2}}\\
-\mathbf{1}_{\frac{N}{2}\times\frac{N}{2}} & 0
\end{array}\right).
\end{align}
The symmetric case has residual symmetry $SO(N)$ and the anti-symmetric
one has $Sp(N)$. In the symmetric case the solution of the equation
(\ref{eq:eqs}) is 
\begin{align}
Y_{i}^{\alpha} & =\bar{Y}_{i},\qquad i=1,\dots,N,\label{eq:ONC1-2-1}
\end{align}
where the bar denotes the complex conjugation. For the anti-symmetric
$\kappa$ the equation (\ref{eq:eqs}) has the solution
\begin{align}
Y_{i}^{\alpha} & =-\bar{Y}_{\frac{N}{2}+i},\qquad i=1,\dots,\frac{N}{2},\\
Y_{\frac{N}{2}+i}^{\alpha} & =\bar{Y}_{i},\qquad i=\frac{N}{2}+1,\dots,N.
\end{align}
We can see that this is not a consistent crosscap because the square
of this transformation is $-1$.

In summary, for the $\mathbb{C}\mathrm{P}^{N-1}$ sigma model the
integrable crosscaps correspond to the residual symmetries $SU(k)\times SU(N-k)\times U(1)$
or $SO(N)$ and the concrete identifications are given by the equations
(\ref{eq:ONC1-2}-\ref{eq:ONC2-2}) or (\ref{eq:ONC1-2-1}). In the
table \ref{tab:CP3} we enumerate these possibilities explicitly for
the $\mathbb{C}\mathrm{P}^{3}$ .

\begin{table}

\caption{Crosscaps of $\mathbb{C}\mathrm{P}^{3}$ sigma model. }

\begin{doublespace}
\begin{centering}
\begin{tabular}{|c|c|}
\hline 
Residual symmetry & $(Y_{1},Y_{2},Y_{3},Y_{4})^{\alpha}$\tabularnewline
\hline 
\hline 
\multirow{2}{*}{$SU(4)$} & $(Y_{1},Y_{2},Y_{3},Y_{4})$\tabularnewline
\cline{2-2} 
 & $(-Y_{1},-Y_{2},-Y_{3},-Y_{4})$\tabularnewline
\hline 
\multirow{2}{*}{$SU(3)\times U(1)$} & $(Y_{1},Y_{2},Y_{3},-Y_{4})$\tabularnewline
\cline{2-2} 
 & $(Y_{1},-Y_{2},-Y_{3},-Y_{4})$\tabularnewline
\hline 
$SU(2)\times SU(2)\times U(1)$ & $(Y_{1},Y_{2},-Y_{3},-Y_{4})$\tabularnewline
\hline 
\hline 
$SO(4)$ & $(-\bar{Y}_{3},-\bar{Y}_{4},\bar{Y}_{1},\bar{Y}_{2})$\tabularnewline
\hline 
\end{tabular}
\par\end{centering}
\end{doublespace}
\label{tab:CP3}
\end{table}

\section{Crosscaps in the AdS/CFT duality\label{sec:Crosscaps-in-AdS/CFT}}

In this section we apply the results of the previous sections to the
classical sigma models which are relevant in the $AdS_{5}/CFT_{4}$
and the $AdS_{4}/CFT_{3}$ dualities. The string theory side we have
type IIB superstrings on the $AdS_{5}\times S^{5}$ and type IIA superstrings
on the $AdS_{4}\times\mathbb{CP}^{3}$. The dual field theories are
the $\mathcal{N}=4$ SYM and the ABJM theories. The isometries of
the $AdS$ and the $S^{5}$ (or $\mathbb{CP}^{3}$) correspond to
the conformal symmetries and the $R$-symmetries of the field theories.
In the following we also use the latter names for the isometries of
the target spaces.

In the following, we classify the integrable crosscaps of classical
sigma models corresponding these superstrings, but we also make a
few physically reasonable restrictions. We concentrate on the half-BPS
crosscaps. Furthermore, we require some breaking of the conformal
symmetry since otherwise they would not be non-vanishing one-point
functions. On the other hand we want the dilation operator to be unbroken,
in other words, the residual symmetry should contain a lower-dimensional
conformal symmetry group.

\subsection{$AdS_{5}/CFT_{4}$}

In this subsection we classify the integrable crosscaps for type IIB
strings on the $AdS_{5}\times S^{5}$. The classical string can be
described as a sigma model on the supercoset \cite{Metsaev:1998it}
\begin{equation}
\frac{PSU(2,2|4)}{SO(1,4)\times SO(5)}.\label{eq:coset}
\end{equation}
We can define the current in the usual way $J=g^{-1}dg$ where $g(\tau,\sigma)\in PSU(2,2|4)$.
The superalgebra $\mathfrak{psu}(2,2|4)$ has a $\mathbb{Z}_{4}$
automorphism for which the current decomposes as $J=J^{(0)}+J^{(1)}+J^{(2)}+J^{(3)}$.
We can define the fixed frame currents as $j^{(m)}=gJ^{(m)}g^{-1}$.
The Lax pair can be written as \cite{Dekel:2011ja}
\begin{equation}
A(\lambda)=(\lambda-1)j^{(1)}+\frac{1}{2}(\lambda-\lambda^{-1})^{2}j^{(2)}+(\lambda^{-1}-1)j^{(3)}-\frac{1}{2}(\lambda^{2}-\lambda^{-2})*j^{(2)}.\label{eq:superLax}
\end{equation}
The theory has global $PSU(2,2|4)$ symmetry which acts on the group
element as $g\to\bar{g}g$ where $\bar{g}\in PSU(2,2|4)$ is a constant
group element. The current $J$ is invariant under this transformation
however the fixed frame currents transform as $j^{(m)}\to\bar{g}j^{(m)}\bar{g}^{-1}$
therefore the Lax connection also transforms as
\begin{equation}
A(\lambda)\to\bar{g}A(\lambda)\bar{g}^{-1}.\label{eq:Glob}
\end{equation}

Let us continue with the crosscap condition (\ref{eq:kL-1}). At first
let $A^{\beta}(\lambda)=A(\lambda)$ therefore
\begin{equation}
A_{\sigma}(\lambda)\Biggr|_{\tau=0,\sigma=\sigma_{0}}=\kappa^{-1}A_{\sigma}(-\lambda)\kappa\Biggr|_{\tau=0,\sigma=\sigma_{0}+\frac{L}{2}}.
\end{equation}
Applying the transformation (\ref{eq:Glob}), we obtain that
\begin{equation}
A_{\sigma}(\lambda)\Biggr|_{\tau=0,\sigma=\sigma_{0}}=\bar{g}^{-1}\kappa^{-1}\bar{g}A_{\sigma}(-\lambda)\bar{g}^{-1}\kappa\bar{g}\Biggr|_{\tau=0,\sigma=\sigma_{0}+\frac{L}{2}}.
\end{equation}
 We can see that this condition breaks the global $PSU(2,2|4)$ symmetry,
the residual symmetry is defined by
\begin{equation}
\bar{g}^{-1}\kappa\bar{g}=\kappa.
\end{equation}
We saw that, the consistency of the crosscap requires that $\kappa^{2}=\mathbf{1}$
therefore the possible $\kappa$-s (up to global rotations) are
\begin{equation}
\kappa=\mathrm{diag}(\underbrace{1,\dots,1}_{k},\underbrace{-1,\dots,-1}_{4-k}|\underbrace{-1,\dots,-1}_{4-l},\underbrace{1,\dots,1}_{l}).
\end{equation}
Let us forget about the signature for a moment. The $\kappa$-matrix
breaks the global symmetry $\mathfrak{gl}(4|4)$ to $\mathfrak{gl}(k|l)\oplus\mathfrak{gl}(4-k|4-l)$
(up to $\mathfrak{u}(1)$ factors). Let us concentrate on the 1/2
BPS configurations. We have three possibilities
\begin{equation}
\mathfrak{gl}(2|2)\oplus\mathfrak{gl}(2|2),\qquad\mathfrak{gl}(2|4)\oplus\mathfrak{gl}(2),\qquad\mathfrak{gl}(4|2)\oplus\mathfrak{gl}(2),\qquad\mathfrak{gl}(3|2)\oplus\mathfrak{gl}(1|2)
\end{equation}
The last possibility contains bosonic subalgebra $\mathfrak{u}(3)$
but we already show that this symmetry has no consistent crosscap
for the sigma models neither on the $S^{5}$ or the $AdS_{5}$. The
$\mathfrak{gl}(4|2)\oplus\mathfrak{gl}(2)$ case preserves the full
conformal symmetry $SO(2,4)$ which means there cannot be non-vanishing
one-point function in the CFT side therefore we neglect this case.
We can see that only the first two possibilities remain.
\begin{itemize}
\item For the $\mathfrak{gl}(2|2)\oplus\mathfrak{gl}(2|2)$ case the conformal
symmetry breaks as $SO(2,4)\to SO(2,2)\times SO(2)$ or $SO(2)\times SO(4)$
and we already derived which are the corresponding crosscap conditions
on the Poincaré patch, see table \ref{tab:AdS5}. The $R$-symmetry
breaks as $SO(6)\to SO(2)\times SO(4)$ and we already derived which
are the corresponding crosscap conditions on the sphere, see table
\ref{tab:S5}. 
\item For the $\mathfrak{gl}(2|4)\oplus\mathfrak{gl}(2)$ case the conformal
symmetry breaks as $SO(2,4)\to SO(2,2)\times SO(2)$ or $SO(2)\times SO(4)$
and the $R$-symmetry is unbroken.
\end{itemize}
Let us continue with the the crosscap conditions (\ref{eq:kL-1})
with the transformation $A^{\beta}(\lambda)=-M^{-1}A(\lambda)^{st}M$
where 
\begin{equation}
M=\left(\begin{array}{ccc}
\mathbf{1}_{4\times4} & 0 & 0\\
0 & 0 & \mathbf{1}_{2\times2}\\
0 & -\mathbf{1}_{2\times2} & 0
\end{array}\right),
\end{equation}
and the super-transposition $\:^{st}$ acts in the usual way
\begin{equation}
\left(\begin{array}{cc}
A & Q\\
S & B
\end{array}\right)^{st}=\left(\begin{array}{cc}
A^{t} & -S^{t}\\
Q^{t} & B^{t}
\end{array}\right),
\end{equation}
where $A,B$ and $Q,S$ are $4\times4$ bosonic and fermionic matrices.
Using these definitions we can show that the $\beta$ is a $\mathbb{Z}_{2}$
automorphism of $\mathfrak{gl}(4|4)$. The crosscap conditions read
as
\begin{equation}
A_{\sigma}(\lambda)\Biggr|_{\tau=0,\sigma=\sigma_{0}}=-\kappa^{-1}M^{-1}A_{\sigma}(-\lambda)^{st}M\kappa\Biggr|_{\tau=0,\sigma=\sigma_{0}+\frac{L}{2}}.
\end{equation}
Applying this transformation twice we obtain that
\begin{equation}
A_{\sigma}(\lambda)=\kappa^{-1}M^{-1}(\kappa^{-1}M^{-1}A_{\sigma}(\lambda)^{st}M\kappa)^{st}M\kappa.
\end{equation}
For bosonic $\kappa$-matrix, we obtain that
\begin{multline}
A_{\sigma}(\lambda)=\kappa^{-1}M^{-1}\kappa^{t}MM^{-1}(M^{-1}A_{\sigma}(\lambda)^{st}M)^{st}MM^{-1}\left(\kappa^{-1}\right)^{t}M\kappa=\\
\left(\kappa^{-1}M^{-1}\kappa^{t}M\right)A_{\sigma}(\lambda)\left(M^{-1}\left(\kappa^{-1}\right)^{t}M\kappa\right),
\end{multline}
where we used the identity
\[
M^{-1}(M^{-1}A_{\sigma}(\lambda)^{st}M)^{st}M=A_{\sigma}(\lambda).
\]
For a consistent crosscap we obtained the following constraint for
the $\kappa$-matrix
\begin{equation}
\kappa=\pm M^{-1}\kappa^{t}M.
\end{equation}
We have two types of $\kappa$-matrices (up to global rotations).
For the $+$ sign the $\kappa$-matrix reads as
\begin{equation}
\kappa=\mathbf{1},\label{eq:ps}
\end{equation}
and for the $-$ sign the $\kappa$-matrix is
\begin{equation}
\kappa=\left(\begin{array}{cccc}
0 & \mathbf{1}_{2\times2} & 0 & 0\\
-\mathbf{1}_{2\times2} & 0 & 0 & 0\\
0 & 0 & 0 & -\mathbf{1}_{2\times2}\\
0 & 0 & \mathbf{1}_{2\times2} & 0
\end{array}\right).\label{eq:ns}
\end{equation}
Applying the transformation (\ref{eq:Glob}), we obtain that
\begin{equation}
A_{\sigma}(\lambda)\Biggr|_{\tau=0,\sigma=\sigma_{0}}=\bar{g}^{-1}\kappa^{-1}M^{-1}\left(\bar{g}^{-1}\right)^{st}A_{\sigma}(-\lambda)^{st}\bar{g}^{st}M\kappa\bar{g}\Biggr|_{\tau=0,\sigma=\sigma_{0}+\frac{L}{2}}.
\end{equation}
We can see that this condition breaks the global $PSU(2,2|4)$ symmetry
and the residual symmetry is defined by
\begin{equation}
\bar{g}^{st}M\kappa\bar{g}=M\kappa.
\end{equation}
We already saw that, for a consistent crosscap we have two types of
$\kappa$-matrices. For the symmetric $\kappa$-matrix (\ref{eq:ps})
we have
\begin{equation}
M\kappa=M=\left(\begin{array}{ccc}
\mathbf{1}_{4\times4} & 0 & 0\\
0 & 0 & \mathbf{1}_{2\times2}\\
0 & -\mathbf{1}_{2\times2} & 0
\end{array}\right)
\end{equation}
for which the residual symmetry is $\mathfrak{osp}(4|4)$. For the
anti-symmetric $\kappa$-matrix (\ref{eq:ps}) we have
\begin{equation}
M\kappa=\left(\begin{array}{ccc}
0 & \mathbf{1}_{2\times2} & 0\\
-\mathbf{1}_{2\times2} & 0 & 0\\
0 & 0 & \mathbf{1}_{4\times4}
\end{array}\right)
\end{equation}
for which the residual symmetry is also $\mathfrak{osp}(4|4)$ but
now the first $4\times4$ bosonic block breaks to $\mathfrak{sp}(4)$
and second breaks to $\mathfrak{so}(4)$. For the sake of clarity,
we denote this residual symmetry as $\mathfrak{spo}(4|4)$. 

We can see that these are 1/2 BPS crosscaps. In summary we have two
possibilities:
\begin{itemize}
\item For the $\mathfrak{osp}(4|4)$ residual symmetry, the conformal symmetry
breaks as $SO(2,4)\to SO(2,1)\times SO(3)$ and $R$-symmetry breaks
as $SO(6)\to SO(5)$.
\item For the $\mathfrak{spo}(4|4)$ residual symmetry, the conformal symmetry
breaks as $SO(2,4)\to SO(2,3)$ and $R$-symmetry breaks as $SO(6)\to SO(3)\times SO(3).$
\end{itemize}
In the table \ref{tab:AdS5CFT4} we summarized the crosscaps for the
sigma model on the coset (\ref{eq:coset}). This table shows the residual
symmetries of the possible crosscaps and the concrete realizations
(which is given by the $\mathbb{Z}_{2}$ automorphism) on the $AdS_{5}$
and the $S^{5}$ are given by the tables \ref{tab:AdS5} and \ref{tab:S5}.

\begin{table}
\caption{Possible crosscaps in the $AdS_{5}/CFT_{4}$}

\begin{doublespace}
\begin{centering}
\begin{tabular}{|c|c|c|}
\hline 
Residual symmetry & Residual isometries on $AdS_{5}$ & Residual isometries on $S^{5}$\tabularnewline
\hline 
\hline 
\multirow{2}{*}{$\mathfrak{gl}(2|2)\oplus\mathfrak{gl}(2|2)$} & $SO(2,2)\times SO(2)$ & \multirow{2}{*}{$SO(4)\times SO(2)$}\tabularnewline
\cline{2-2} 
 & $SO(2)\times SO(4)$ & \tabularnewline
\hline 
\multirow{2}{*}{$\mathfrak{gl}(2|4)\oplus\mathfrak{gl}(2)$} & $SO(2,2)\times SO(2)$ & \multirow{2}{*}{$SO(6)$}\tabularnewline
\cline{2-2} 
 & $SO(2)\times SO(4)$ & \tabularnewline
\hline 
\hline 
\multirow{2}{*}{$\mathfrak{osp}(4|4)$} & $SO(2,1)\times SO(3)$ & \multirow{1}{*}{$SO(5)$}\tabularnewline
\cline{2-3} \cline{3-3} 
 & $SO(2,3)$ & \multirow{1}{*}{$SO(3)\times SO(3)$}\tabularnewline
\hline 
\end{tabular}
\par\end{centering}
\end{doublespace}
\label{tab:AdS5CFT4}
\end{table}

\begin{table}
\caption{Possible crosscaps in the $AdS_{4}/CFT_{3}$}

\begin{doublespace}
\begin{centering}
\begin{tabular}{|c|c|c|}
\hline 
Residual symmetry & Residual isometries of $AdS_{4}$ & Residual isometries of $\mathbb{CP}^{3}$\tabularnewline
\hline 
\hline 
\multirow{1}{*}{$\mathfrak{osp}(3|2)\oplus\mathfrak{osp}(3|2)$} & $SO(2,2)$ & \multirow{1}{*}{$SO(4)$}\tabularnewline
\hline 
\multirow{1}{*}{$\mathfrak{osp}(6|2)\oplus\mathfrak{sp}(2)$} & $SO(2,2)$ & \multirow{1}{*}{$SU(4)$}\tabularnewline
\hline 
\multirow{1}{*}{$\mathfrak{osp}(4|2)\oplus\mathfrak{osp}(2|2)$} & $SO(2,2)$ & \multirow{1}{*}{$SU(2)\times SU(2)\times U(1)$}\tabularnewline
\hline 
\hline 
\multirow{1}{*}{$\mathfrak{gl}(3|2)$} & $SO(2,1)\times SO(2)$ & \multirow{1}{*}{$SU(3)\times U(1)$}\tabularnewline
\hline 
\end{tabular}
\par\end{centering}
\end{doublespace}
\label{tab:AdS4CFT3}
\end{table}

\subsection{$AdS_{4}/CFT_{3}$}

In this subsection we classify the integrable crosscaps for type IIA
superstrings on $AdS_{4}\times\mathbb{CP}^{3}$. The classical string
can be described as a sigma model on the supercoset \cite{Arutyunov:2008if}
\begin{equation}
\frac{OSP(6|4)}{SO(1,3)\times U(3)}.\label{eq:coset2}
\end{equation}
We can define the current in the usual way $J=g^{-1}dg$ where $g(\tau,\sigma)\in OSP(6|4)$.
We can also define the fixed frame currents and Lax connect operators
in the same way as before (\ref{eq:superLax}). We saw that the consistency
of the crosscap requires that $\kappa^{2}=\mathbf{1}$ and now the
current has the symmetry $J=-VJ^{st}V^{-1}$where
\begin{equation}
V=\left(\begin{array}{ccc}
\mathbf{1}_{6\times6} & 0 & 0\\
0 & 0 & \mathbf{1}_{2\times2}\\
0 & -\mathbf{1}_{2\times2} & 0
\end{array}\right),
\end{equation}
therefore the consistency also requires that $\kappa^{t}=\pm V\kappa V^{-1}$
(the $\kappa$-matrix is bosonic). The theory has global $OSP(6|4)$
symmetry which act on the group element as $g\to\bar{g}g$ where $\bar{g}\in OSP(6|4)$
is a constant group element. Repeating the argument of the previous
subsection we obtain that the residual symmetry is defined by
\begin{equation}
\bar{g}^{-1}\kappa\bar{g}=\kappa.
\end{equation}

The type $\kappa^{t}=+V\kappa V^{-1}$ matrices can be written (up
to global rotations) as 
\begin{equation}
\kappa=\mathrm{diag}(\underbrace{1,\dots,1}_{k},\underbrace{-1,\dots,-1}_{6-k}|\underbrace{-1,\dots,-1}_{\frac{l}{2}},\underbrace{1,\dots,1}_{\frac{4-l}{2}},\underbrace{-1,\dots,-1}_{\frac{l}{2}},\underbrace{1,\dots,1}_{\frac{4-l}{2}}).
\end{equation}
This $\kappa$-matrix breaks the global symmetry $\mathfrak{osp}(6|4)$
to $\mathfrak{osp}(k|l)\oplus\mathfrak{osp}(6|4-l)$. Let us concentrate
on the 1/2 BPS configurations. We have five possibilities
\[
\mathfrak{osp}(3|2)\oplus\mathfrak{osp}(3|2),\quad\mathfrak{osp}(6|2)\oplus\mathfrak{sp}(2),\quad\mathfrak{osp}(3|4)\oplus\mathfrak{so}(3),\quad\mathfrak{osp}(4|2)\oplus\mathfrak{osp}(2|2),\quad\mathfrak{osp}(5|2)\oplus\mathfrak{osp}(1|2).
\]
In the last possibility the $R$-symmetry breaks as $\mathfrak{su}(4)\to\mathfrak{so}(5)\cong\mathfrak{sp}(4)$
but we already showed that this symmetry has no consistent crosscap
for the sigma model on $\mathbb{CP}^{3}$. The $\mathfrak{osp}(3|4)\oplus\mathfrak{so}(3)$
case preserves the full conformal symmetry $SO(2,3)$ which means
there cannot be non-vanishing one-point function in the CFT side therefore
we neglect this case. We can see that three possibilities remain.
\begin{itemize}
\item For the $\mathfrak{osp}(3|2)\oplus\mathfrak{osp}(3|2)$ case the conformal
symmetry breaks as $SO(2,3)\to SO(2,2)$ and we already derived which
are the corresponding crosscap conditions on the Poincaré patch, see
table \ref{tab:AdS4}. The $R$-symmetry breaks as $SU(4)\to SO(4)$
and we already derived which are the corresponding crosscap conditions
on the $\mathbb{CP}^{3}$, see table \ref{tab:CP3}. 
\item For the $\mathfrak{osp}(6|2)\oplus\mathfrak{sp}(2)$ case the conformal
symmetry breaks as $SO(2,3)\to SO(2,2)$ and the $R$-symmetry is
preserved. 
\item For the $\mathfrak{osp}(4|2)\oplus\mathfrak{osp}(2|2)$ case the conformal
symmetry breaks as $SO(2,3)\to SO(2,2)$ and the $R$-symmetry breaks
as $SU(4)\to SU(2)\times SU(2)\times U(1)$. 
\end{itemize}
The type $\kappa^{t}=-V\kappa V^{-1}$ matrices can be written (up
to global rotations) as 
\begin{equation}
\kappa=\left(\begin{array}{cccc}
0 & \mathbf{1}_{3\times3} & 0 & 0\\
-\mathbf{1}_{3\times3} & 0 & 0 & 0\\
0 & 0 & \mathbf{1}_{2\times2} & 0\\
0 & 0 & 0 & -\mathbf{1}_{2\times2}
\end{array}\right).
\end{equation}
This $\kappa$-matrix breaks the global symmetry $\mathfrak{osp}(6|4)$
to $\mathfrak{gl}(3|2)$ which is 1/2 BPS. The conformal symmetry
breaks as $SO(2,3)\to SO(2,1)\times SO(2)$ and the $R$-symmetry
breaks as $SU(4)\to SU(3)\times U(1)$. 

In the table \ref{tab:AdS4CFT3} we summarized the crosscaps for the
sigma model on the coset (\ref{eq:coset2}). This table shows the
residual symmetries of the possible crosscaps and the concrete realizations
(which is given by the $\mathbb{Z}_{2}$ automorphism) on the $AdS_{4}$
and the $\mathbb{CP}^{3}$ are given by the tables \ref{tab:AdS4}
and \ref{tab:CP3}.

\subsection{Crosscaps at weak coupling}

In the literature there is an other classification of crosscaps which
is relevant in the field theory side \cite{Gombor:2022deb}. This
paper contains the classifications of crosscap states of $SO(6)$
and the alternating $SU(4)$ spin chains which describes the $\mathcal{N}=4$
SYM and ABJM theories at weak coupling. In this paper we used such
parametrization of the bosonic spaces which can be directly match
to the CFT side of the duality. The $S^{5}$ and $\mathbb{C}\mathrm{P}^{3}$
are parameterized with the coordinates $(\phi_{1},\phi_{2},\dots,\phi_{6})$
and $(Y_{1},Y_{2},Y_{3},Y_{4})$ which corresponds to the scalar field
of the $\mathcal{N}=4$ SYM and ABJM theories. The corresponding $\mathbb{Z}_{2}$
isometries can be found in the tables \ref{tab:S5} and \ref{tab:CP3}.
We can compare the proposed crosscaps at strong coupling (result of
this paper) and at weak coupling (result of \cite{Gombor:2022deb})
and we find that they are almost the same. There are two differences.
There is no $U(3)$ symmetric crosscap for the $S^{5}$ sigma model
and there is no $Sp(4)$ symmetric crosscap for the $\mathbb{C}\mathrm{P}^{3}$
sigma model. 

The spin chain classifications belong only to the scalar sectors and
they tell nothing about the effect of the crosscap on the spacetime.
The main advantage of the analysis of this paper is the following.
The classification of the crosscaps of the sigma models on the supercosets
tells us what are the consistent combinations of the crosscaps on
the $AdS_{5}$ (or $AdS_{4}$) and the $S^{5}$ (or $\mathbb{C}\mathrm{P}^{3}$),
see tables \ref{tab:AdS5CFT4} and \ref{tab:AdS4CFT3}. Therefore
we obtained candidates for the crosscaps on the $CFT$ side of the
duality. In the tables \ref{tab:AdS5} and \ref{tab:AdS4} we can
find the possible crosscaps of $AdS_{5}$ and $AdS_{4}$ which are
parameterized with the Poincaré coordinates $(z,x_{0},x_{1},\dots)$
therefore we can obtain the identifications of the spacetime of the
$\mathcal{N}=4$ SYM and ABJM by taking the limit $z=0$. 

In summary we obtained propositions for the possible integrable crosscaps
of $\mathcal{N}=4$ SYM and ABJM theories. The residual symmetries
are listed in the tables \ref{tab:AdS5CFT4} and \ref{tab:AdS4CFT3}.
Each symmetry classes define identifications on the spacetime and
the scalar fields. For each symmetry classes the corresponding identifications
of the spacetime and scalar fields can be found in the tables \ref{tab:S5}-\ref{tab:CP3}
(take the $z=0$ limit). We emphasize that this classification is
a conjecture for the $\mathcal{N}=4$ SYM and ABJM theories and this
paper is not intended to provide a concrete field theory description
(if that is even possible).

\section{Conclusion}

In this paper we generalized the Lax description of the sigma models
with boundaries in time. For the usual local boundary conditions we
obtained constraints for the classical monodromy and transfer matrices
and these are the classical analogs of the $KT$-relations \cite{Gombor:2021hmj}
and the integrability conditions \cite{Piroli:2017sei} of the quantum
theories. We also generalized this framework for the crosscaps where
the classical analogs of the $KT$-relations \cite{Gombor:2022deb}
and the integrability conditions are also appeared.

Based on this framework, we classified the integrable crosscaps for
the sigma models with target spaces $SU(N)$, $S^{N-1}$, $AdS_{N-1}$
and $\mathbb{C}\mathrm{P}^{N-1}$. The classification is based on
the possible residual symmetries of the crosscaps. We gave the defining
isometries of the crosscaps for each residual symmetry classes. We
also investigated the supercosets which are relevant in the $AdS_{5}/CFT_{4}$
and the $AdS_{4}/CFT_{3}$ dualities. We classified the integrable
1/2 BPS crosscaps based on the residual symmetries (see tables \ref{tab:AdS5CFT4}
and \ref{tab:AdS4CFT3}). The corresponding defining isometries of
the bosonic subspaces can be found in the tables \ref{tab:S5}-\ref{tab:CP3}.
It is important to emphasize that this classification only applies
to classical sigma models, and at this point we do not know that which
versions have consistent holographic descriptions. This is an interesting
question for a future research.

\section*{Acknowledgments}

I thank Balázs Pozsgay, Zoltán Bajnok and Georgios Linardopoulos for
the useful discussions and the NKFIH grant K134946 for support.

\bibliographystyle{elsarticle-num}
\bibliography{refs}

\end{document}